\documentclass[sigconf,screen,authorversion]{acmart}

\usepackage[utf8]{inputenc}        
\usepackage[nameinlink]{cleveref}  

\newcommand{\enquote}[1]{``#1''}  

\newcommand{\ttxplatform}{INJECT Exercise Platform}
\newcommand{\ttxplatformshort}{IXP}

\copyrightyear{2024}
\acmYear{2024}
\setcopyright{rightsretained}
\acmConference[ITiCSE 2024]{Proceedings of the 2024 Innovation and Technology in Computer Science Education V. 1}{July 8--10, 2024}{Milan, Italy}
\acmBooktitle{Proceedings of the 2024 Innovation and Technology in Computer Science Education V. 1 (ITiCSE 2024), July 8--10, 2024, Milan, Italy}
\acmDOI{10.1145/3649217.3653639}
\acmISBN{979-8-4007-0600-4/24/07}

\begin{document}

\title{From Paper to Platform: Evolution of a Novel~Learning~Environment for Tabletop Exercises}

\author{Valdemar Švábenský}
\orcid{0000-0001-8546-280X}
\affiliation{
    \institution{Masaryk University}
    \department{Faculty of Informatics}
    \city{Brno}
    \country{Czech Republic}
}
\email{valdemar@mail.muni.cz}

\author{Jan Vykopal}
\orcid{0000-0002-3425-0951}
\affiliation{
    \institution{Masaryk University}
    \department{Faculty of Informatics}
    \city{Brno}
    \country{Czech Republic}
}
\email{vykopal@fi.muni.cz}

\author{Martin Horák}
\orcid{0000-0002-1835-6465}
\affiliation{
    \institution{Masaryk University}
    \department{Faculty of Informatics}
    \city{Brno}
    \country{Czech Republic}
}
\email{horak.martin@fi.muni.cz}

\author{Martin Hofbauer}
\orcid{0009-0005-3998-9164}
\affiliation{
    \institution{Masaryk University}
    \department{Faculty of Informatics}
    \city{Brno}
    \country{Czech Republic}
}
\email{hofbauer@fi.muni.cz}

\author{Pavel Čeleda}
\orcid{0000-0002-3338-2856}
\affiliation{
    \institution{Masaryk University}
    \department{Faculty of Informatics}
    \city{Brno}
    \country{Czech Republic}
}
\email{celeda@fi.muni.cz}

\begin{abstract}
For undergraduate students of computing, learning to solve complex practical problems in a team is an essential skill for their future careers. This skill is needed in various fields, such as in cybersecurity and IT governance. Tabletop exercises are an innovative teaching method used in practice for training teams in incident response and evaluation of contingency plans. However, tabletop exercises are not yet widely established in university education. This paper presents data and teaching experience from a cybersecurity course that introduces tabletop exercises in classrooms using a novel technology: \ttxplatform\ (\ttxplatformshort), a web-based learning environment for delivering and evaluating the exercises. This technology substantially improves the prior practice, since tabletop exercises worldwide have usually been conducted using pen and paper. Unlike in traditional tabletop exercises, which are difficult to evaluate manually, \ttxplatformshort\ provides insights into students' behavior and learning based on automated analysis of interaction data. We demonstrate \ttxplatformshort's capabilities and evolution by comparing exercise sessions hosted throughout three years at different stages of the platform's readiness. The analysis of student data is supplemented by the discussion of the lessons learned from employing \ttxplatformshort\ in computing education contexts. The data analytics enabled a detailed comparison of the teams' performance and behavior. Instructors who consider innovating their classes with tabletop exercises may use \ttxplatformshort\ and benefit from the insights in this paper.
\end{abstract}



\maketitle

\section{Introduction}

\textit{Collaborative problem-solving of complex issues} (CPSCI), such as the resolution of incidents in large organizations that critically rely on information technology (IT), is a core competency for the twenty-first-century workforce~\cite{Fiore2018, cc2020}. However, many university graduates lack necessary skills in these areas~\cite{Fiore2018}.

University students of applied computing (a target student demographic of this paper) learn CPSCI in cybersecurity and IT governance courses, among others. These courses cover topics like cyber incident response, emergency readiness, information sharing, or contingency plan validation when managing an IT infrastructure.

Computing educators found it difficult to provide students with practical learning experience in such courses~\cite{Ottis2014}. Thus, researchers and instructors have been exploring innovative ways to teach these interdisciplinary topics, which connect technological and human aspects of computing, in an immersive and meaningful way~\cite{Ottis2014}.

\subsection{What Are Tabletop Exercises?}

A \textit{tabletop exercise} (TTX) is a type of a teaching activity designed to train professional teams in incident response to a crisis situation~\cite{NIST2006}. The simulated crisis happens in the context of business operations in an organization, e.g., a phishing attack on employees or malware infecting the company infrastructure. The team members (exercise participants) hold various roles in the organization, e.g., manager or cybersecurity incident responder~\cite{Angafor2023}. During the exercise, the team discusses which actions to take to effectively respond to the emergency while following proper protocols and regulations. The team discussions are facilitated by instructors, who also present an exercise debriefing at the end.

TTXs are an effective educational tool, which enhance incident preparedness of individuals, particularly their communication, coordination, and collaboration~\cite{Angafor2023}. In computing education, TTXs provide students with realistic incident response experience and deepen their understanding of related processes. TTXs are especially relevant for cybersecurity and information security management courses and align with broader IT governance courses~\cite{ENISA2010}.

\subsection{Problem Statement and Innovation}

TTXs differ from technical hands-on exercises in an emulated computer infrastructure, such as in a cyber range~\cite{yamin2020} or a locally virtualized learning environment~\cite{Vykopal2021scalable}. Instead, TTXs are much more lightweight and do not dive into technical matters deeply. They are traditionally conducted using pen and paper or simple online office applications, such as Google Forms, to collect participant responses. The advantage of this approach is its low cost and low barrier to entry. On the other hand, the assessment of the participating teams has to be done manually by the instructors, which is highly time-consuming. It takes days or even weeks until the trainees can receive educational feedback, which diminishes its effectiveness and decreases learning gains from the TTX.

We aim to transition the TTX format from this low-tech approach into \ttxplatform\ (\ttxplatformshort): a novel, lightweight, open-source environment for supporting the deployment and evaluation of TTXs. This represents a major innovation that automates repetitive tasks for instructors, leaving them more room for teaching. Since \ttxplatformshort\ automatically collects exercise data, it can deliver pedagogical insights and feedback using the methods of learning analytics. This paper shares our experience in deploying \ttxplatformshort\ in computing classes and analyzes student data from these classes.

\subsection{Goals and Scope of This Paper}
\label{subsec:rq}

We developed a novel TTX, which we deployed on three occasions (\enquote{runs}) with three groups of learners. In the first run, we used only online Microsoft Office~\cite{msoffice} applications. In the second run, we used a simple prototype of the TTX platform. Finally, the third run demonstrated a more developed version of \ttxplatformshort. Our goal is to compare the student data and teacher perspective on facilitating the TTX in these three different versions of the learning environment. Specifically, this paper explores the following research questions:
\begin{enumerate}
    \item \textit{What types of insights about the student behavior and learning can the platform deliver to instructors?}
    \item \textit{What is the instructors' teaching experience when comparing the three exercise runs?}
\end{enumerate}

\section{Related Work}
\label{sec:related-work}

As this paper focuses on transitioning TTXs from pen-and-paper format to a software platform, we reviewed literature covering all three stages of the transition: in-person pen-and-paper format, online pen-and-paper format, and online platform for TTX delivery. We also review analytics of data from TTXs. Based on our review of related work, we summarize the unique contributions of our paper. 

\subsection{Pen-and-Paper Tabletop Exercises}

\citet{Ottis2014} described how to create lightweight TTXs for cybersecurity education. The TTX detailed in the paper is in-person with two groups of participants: \enquote{red} and \enquote{blue}. Red teams are in charge of creating the exercise \textit{scenario} from the attackers' perspective. (A~TTX scenario is an outline of the sequence of events that drive the exercise and guide participant discussion~\cite{cisa-methodology}.) Blue teams are responsible for handling the attack events. The paper presents observations from eight such TTXs with 250 students in total.

\citet{Angafor2023} used Microsoft Teams to conduct an online pen-and-paper-like TTX for 20 participants from an unnamed company. After the exercise, the participants answered a survey about their awareness of attack mitigation controls, as well as feedback on the completeness of controls currently used in the company. The authors used descriptive statistics to analyze the survey. However, the publication does not include any learning analytics. 

\citet{brilingaité2017} conducted a TTX with students of IT and social sciences. A custom software that hosted the exercise also logged data about user actions. These actions include the number of messages and the number of exercise events in different states of progress, which do not offer extensive possibilities for analysis. The option of further analysis of exercise logs is mentioned, but neither these logs nor the analysis are available.

\subsection{Software for Tabletop Exercises}

TTXs in the cybersecurity context are quite prominent~\cite{enisa-survey1,enisa-survey2}, which leads to research and development of software solutions. While there are companies, such as Privasec~\cite{Privasec} or Red Goat~\cite{redgoat}, that provide paid software for TTXs, open-source solutions exist as well.

We discovered 16 open-source projects for TTXs~\cite{Vykopal2024research} on GitHub, out of which 11 contained software solutions. Most of them are simple and specifically tailored for delivering just one specific exercise scenario. An example is an application~\cite{cli-ttx} that presented the scenario in a few sentences via a command-line interface and asked the participants to discuss the possible solutions. 

While some software solutions are more advanced than just presentations of scenarios, they focus on the cybersecurity aspects of the exercise, as opposed to more discussion-based problems. These solutions are also more technical. An example is Ransomware Simulator~\cite{ransomware-simulator}, which works as a reporting tool for incidents, asking for the event ID, owner, summary, and response to the event. The instructors can add the option of a simulated ransomware attack launched at a specified time, locking participants out of the tool.

The open-source software solution that we consider to have the most features is OpenEx~\cite{openex}. This solution is not specifically tailored for cybersecurity TTXs, and it allows to create and execute different scenarios. Unlike other software we found, OpenEx records logs of participant interactions within the scenario. This enables analyzing the data gathered during the exercise; however, OpenEx does not implement such analysis. Another downside of OpenEx is using real email infrastructure for participant communication, which can lead to delays due to antispam or system outages.

\subsection{Data Analytics in Tabletop Exercises}

As of October 2023, searching \enquote{tabletop exercise} in the multi-faceted citation database Scopus~\cite{scopus} returned 418 papers. While this amount is non-negligible, the number of publications with learning analytics of data from TTXs is low.

\citet{Mare2023} conducted a TTX for 33 experts in cybersecurity and related fields, such as law. The authors analyzed data from two surveys about the participants' behavior, performance, and workload handling. The first survey was carried out immediately after the exercise, and the following survey two weeks later. Higher performance of participants was significantly correlated with their lower levels of perceived stress ($r = 0.30$ to $0.37$, $p = 0.039$). However, this study does not include data about the participants' learning.

\citet{Hsieh2023} compared fire safety knowledge acquisition between drill-based and game-based learning. Although this study did not use the TTX format as defined above, the game-based learning was carried out as a tabletop game. The authors used t-tests to measure the knowledge gain of both groups. The knowledge gain of the game-based group ($t = 12.58$) was substantially larger than for the other group ($t = 6.14$), with $p < 0.001$ for both statistics. 

We did not find any other publication containing learning analytics of TTX data. The only data currently being collected from TTXs are not focused on the educational process, but on feedback on the exercise itself and its perceived usefulness for the participants~\cite{Ottis2014, Kopustinskas2020, yuitaka2022}. These data are valuable for the exercise creators but do not provide deeper insight into the TTX participant behavior.

\subsection{Novel Contributions of This Paper}
\label{subsec:related-work-novelty}

TTXs are suited for computing education, and some software solutions for conducting TTXs exist. However, the existing research does not focus on TTX participant learning behavior. The software solutions do not implement analytics, other than descriptive. So, our work contributes to educators and researchers with these inputs:

\begin{itemize}
    \item We propose an innovated TTX format (\Cref{subsec:exercise-format}).
    \item Since TTXs rarely use dedicated software for evaluation and in-depth analysis, we develop a new learning environment that provides these functionalities (\Cref{subsec:exercise-platform}).
    \item Unlike traditional TTXs, in which the instructors analyze the exercise data manually, we demonstrate the platform's capabilities in automated data collection and analysis. The data come from three runs of a novel TTX deployed in an authentic teaching context (\Cref{sec:methods}).
    \item For instructors and practitioners, we share the practical lessons learned from using the platform (\Cref{sec:results}).
    \item Artifacts associated with this work are available (\Cref{subsec:conclusion-materials}).
\end{itemize}

\section{Tabletop Exercise Delivery}
\label{sec:exercise}

We now define the key features of TTXs. The purpose of this section is twofold: (1) to provide the background for our research study, which is detailed in \Cref{sec:methods}, and (2) to represent a contribution on its own by defining the innovated TTX format and its properties.

\subsection{Proposed Exercise Format}
\label{subsec:exercise-format}

\subsubsection{Participant Roles}

Human participants in a TTX can have one of three roles. \textit{Designers} prepare the exercise and its scenario. \textit{Instructors} facilitate the exercise by guiding the participants and evaluate the exercise at its end. They may or may not be different from designers. \textit{Trainees} attend the exercise to improve their skills. Trainees are grouped into teams that are independent of each other (i.e., each team completes the same tasks in parallel). Each person may have a different role in the team.

\subsubsection{Components of the Exercise}

An \textit{inject} is a pre-scripted message, such as an email, provided to trainees during the TTX. Its purpose is to move the scenario forward and prompt additional actions. For example, it can inform the trainees about a data breach in their company, requiring them to respond accordingly~\cite{NIST2006}.

A \textit{tool} is a simplified simulated version of a real-world computer application/service. Its purpose is to allow trainees to perform actions to respond to injects. For example, trainees in a TTX do not use an actual email client, but an \enquote{email} tool to send and receive in-exercise messages. Another example is a \enquote{browser}, a tool that returns an in-exercise website based on the provided exercise URL.

A \textit{milestone} is a true/false condition that denotes whether a specific team reached a certain important situation in the TTX scenario. Its purpose is to track each team's progress through the TTX. For example, it can mark that a team used an email tool to respond to a query from a manager. The exercise milestones can be completed in any order, and there is rarely a single correct solution. These properties make the team assessment challenging.

\subsubsection{Exercise Workflow}

The exercise is driven by injects, which are either provided by the instructor, or triggered automatically based on time (e.g., an inject is sent after 20 minutes of the exercise) or based on milestone (in)completion. Some injects may be provided to all teams at once; others are conditioned by the team's progress. This allows each team to progress through the scenario independently of other teams. As a result, each team can progress at their own pace, which removes the need for them to wait idle.

When a new inject arrives, trainees in a team discuss the situation to agree on which action to take (e.g., which tool to use). Effective communication under time pressure is a crucial component of TTXs. Trainees do not choose an inject response from pre-defined options, but have to think of a unique open-ended solution. If a team gets stuck, the instructor should help them by asking guiding questions or providing a gentle hint via responding to the team's emails.

\subsection{Exercise Platform}
\label{subsec:exercise-platform}

\begin{figure*}[!t]
\centering
\includegraphics[width=0.85\linewidth]{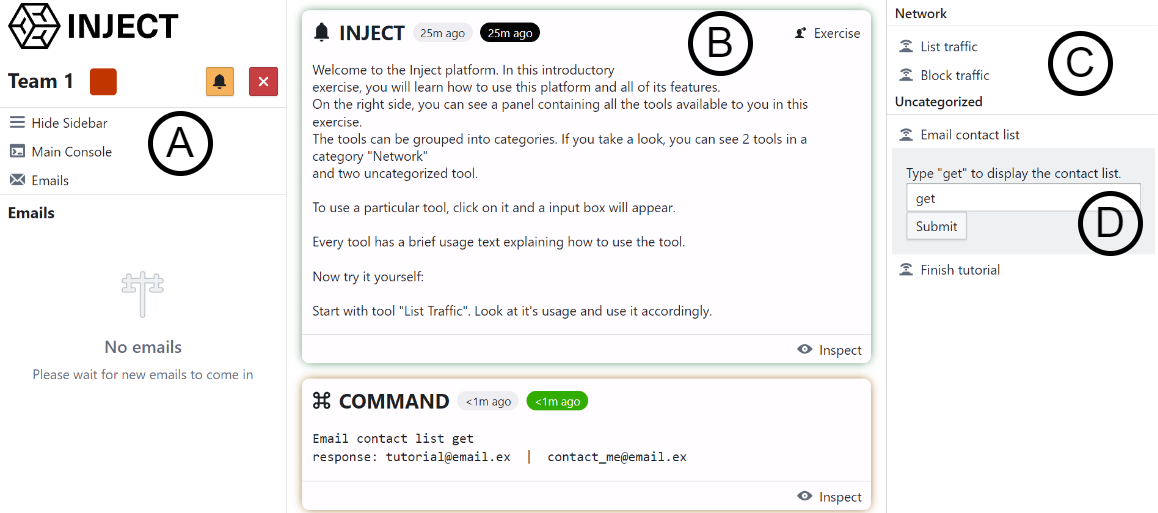}
\caption{Trainees' view of the \ttxplatform. In the left sidebar (\textbf{A}), trainees see injects or email conversations. The middle pane (\textbf{B}) shows injects, emails, and outcomes of tools, depending on the view chosen in the left bar. In the right sidebar (\textbf{C}), trainees see all available tools. After clicking on the tool, a dropdown menu (\textbf{D}) for the tool's arguments appears.}
\Description{Screenshot of the frontend of the Tabletop Exercise Platform.}
\label{fig:frontend}
\end{figure*}

We developed a novel learning environment called \ttxplatform\ (\ttxplatformshort), which is an interactive web application for supporting the delivery and evaluation of TTXs.

Designers can use the platform to instantiate an \textit{exercise definition}, which prescribes the exercise story, injects, available tools, and milestones. An exercise definition is implemented as a set of structured text-based files (in YAML format) that are both human- and machine-readable. It automates a substantial portion of the TTX, since it provides trainees with tools and inject response templates. 

Instructors deploy an exercise definition in \ttxplatformshort\ when they want to host an exercise. The definition is created only once, but thanks to the platform's automation capabilities, it can be deployed repeatedly under the same conditions for different trainees. Compared to manually-hosted exercises, \ttxplatformshort\ significantly reduces the workload and personnel requirements for TTX delivery.

Trainees interact with the scenario through automated tools during the exercise. This interaction moves the scenario forward and impacts the simulated environment. \Cref{fig:frontend} shows the trainees' view of the \ttxplatformshort\ to demonstrate some of the interaction options.

\subsection{Exercise Content Example}

To illustrate the exercise format and the components of the platform, we now describe an exercise that we developed for the \ttxplatformshort. This TTX was also selected for our research study described in \Cref{sec:methods}.

\subsubsection{Learning Objectives}

The TTX was based on a real cyber attack that happened at our university, in order to provide the trainees with an authentic learning experience. The learning objectives of the TTX are: (a) to perform cyber incident triage, (b) to coordinate and execute incident response, and (c) to mitigate the impacts of an incident on a large organization with multiple involved parties.

\subsubsection{Description of the Story}

In the TTX scenario, the trainees assume the role of the members of a \textit{Computer Security Incident Response Team} (CSIRT) of the university. At the beginning, the trainees receive an initial inject: a report of a phishing email targeting the university employees. Several employees have fallen victims to the attack and submitted their login credentials to a fraudulent phishing website. As a result, a malicious actor accessed sensitive information on the employees' internal project server and also deleted important files, making the project website unavailable. As the time progresses, the trainees receive more and more injects in the form of emails from the affected employees, asking the team to take swift action in response to the ongoing emergency.

\subsubsection{Available Tools}

The trainees can take numerous actions at each point of the TTX. They can apply technical solutions (e.g., inspect network data or block traffic to/from certain IP addresses) as well as take managerial/governance steps (e.g., notify the responsible persons according to the data protection law). Each action (e.g.,~using a specific tool) or inaction (e.g., not responding to a query quickly enough) may trigger another inject to propel the scenario forward and place the trainees in the midst of another time-critical issue. To support discussion in a team, only one person from the team can interact with the tools in \ttxplatformshort\ at any given time, after the members consult and mutually agree on their progress.

\section{Research Methods}
\label{sec:methods}

We delivered the exercise described in \Cref{sec:exercise} on three occasions, throughout three years with the total of 91 university students of computing. This section describes the design of our research study. The research questions were posed in \Cref{subsec:rq}.

\subsection{Course Context}

The TTX is the culmination of the course titled \textit{Cybersecurity in an Organization}, which is taught at the Faculty of Informatics, Masaryk University: a large, public university in Central Europe. 

\subsubsection{Learning Outcomes}

The course graduates should understand the role and services of a CSIRT in an organization. Specifically, the course covers knowledge and skills required for the work role of \textit{Cyber Defense Incident Responder} as defined by the NICE Cybersecurity Workforce Framework~\cite{NICE}.

\subsubsection{Teaching Format}

The course is offered once per academic year and spans a standard 13-week Fall semester. It is taught in-person using a combination of flipped classroom sessions, discussion, and homework assignments. The TTX at the course's end provides a hands-on learning experience with the knowledge and skills studied throughout the semester. All teaching materials are written in English, but the language of instruction is local (Czech).

\subsubsection{Student Population}

All students enrolled in the course were students of Faculty of Informatics, and the vast majority pursued their degree in cybersecurity. The class size was up to 42 students. Most students were undergraduates. In the latest semester, we had 34 bachelor-level students (31 of which in the cybersecurity degree program) and 7 master-level students.

\subsection{Field Studies Setup and Participants}

This paper analyzes data and experience from three groups of trainees who completed the TTX. \Cref{tab:ttx-runs} summarizes the three training sessions as the \ttxplatformshort\ readiness increased over three years.

\begin{table}[!ht]
\caption{Information about the three TTX runs.}
\label{tab:ttx-runs}

\begin{tabular}{llll}
Run & Date & Students (team division) & Platform \\ \hline

\#1 & Nov 25, 2021 & 19 (5 teams of 3--4 people) & Documents \\
\#2 & Nov 23, 2022 & 36 (9 teams of 4 people)    & Prototype \\
\#3 & Nov 22, 2023 & 36 (12 teams of 3 people)   & \ttxplatformshort
\end{tabular}

\end{table}

To ensure a fair comparison, we attempted to keep the three TTX runs as consistent and similar as possible. All three runs took place within the same course at the same stage of the semester, on the same exercise, and with the same core instructors (though with slightly different teaching assistants). The modality of all runs was fully in-person. The duration of the exercise was 80--90 minutes.

Before each exercise run, the TTX was thoroughly tested during a dry run with our colleagues and senior graduate students (not students of the course). The purpose of the test run was to verify that the platform is ready for practical usage and that the exercise can be meaningfully completed, and to fix any errors or issues that could negatively impact the learning experience of trainees.

The only substantial difference between the three runs is the subject of examination in this paper -- the TTX platform readiness.
\begin{itemize}
    \item Run 1 was an imitation of a pen-and-paper exercise using shared text-based documents on Microsoft SharePoint.
    \item Run 2 was deployed in the first prototype of the dedicated \ttxplatformshort\ developed as a master's thesis~\cite{Urban2023thesis}.
    \item Run 3 featured the latest version of \ttxplatformshort, which was substantially improved by a dedicated development team, mainly including bug fixes and a better user experience.
\end{itemize}

\subsection{Research Ethics and Data Privacy}

This research did not require an approval from the university's institutional review board. All trainees receive their course points simply for active participation. \ttxplatformshort\ does not store any personal information that could reveal the trainees' identity. The data exported for analysis are anonymous and cannot be linked to specific individuals. The trainees were informed that their anonymized exercise activity data may be used for educational research purposes. Lastly, post-exercise surveys were voluntary and anonymous.

\subsection{Exercise Data Collection}

Since the Run 2, \ttxplatformshort\ provides transparent, automated collection of exercise metadata and actions of trainees. The log records are stored in the standard JSONL format~\cite{jsonl}, and each record has a uniform timestamp with microsecond precision. Per each team, \ttxplatformshort\ gathers and categorizes the records into four log files:
\begin{itemize}
    \item Which injects did the team receive {\scriptsize (\texttt{inject\_categories.jsonl})}.
    \item In-exercise email communication {\scriptsize (\texttt{emails.jsonl})}.
    \item Actions performed using the in-exercise tools {\scriptsize (\texttt{action\_logs.jsonl})}.
    \item Reached milestones {\scriptsize (\texttt{milestones.jsonl})}.
\end{itemize}

The format of the logs was improved for the latest version of \ttxplatformshort\ deployed for Run 3. To enable uniform data analysis, we automatically converted the logs from Run 2 to this new format.

\subsection{Exercise Data Analysis}

To analyze the exercise data, we used a combination of a learning analytics dashboard built into \ttxplatformshort\ and dedicated Python scripts. The scripts process the logs after the TTX ends in order to provide additional analytics for assessing the team performance more deeply, such as correct/incorrect tool usage. The scripts also enable to evaluate the TTX as a whole, by looking at metrics like time needed to reach individual milestones.


\section{Results and Their Discussion}
\label{sec:results}

We present and compare the results of data analyses from Run 2 and 3. Since Run 1 was executed using text-based documents, it did not yield logs in the above-described format. All observations are tied to implications for computing educators.

\subsection{Analysis of the TTX Data of Trainees}

\subsubsection{Run 2}

The TTX had 14 defined milestones, which captured actions such as blocking traffic from certain sources, communicating with the affected users, and notifying responsible parties. The teams reached between 5 and 12 milestones, with an average of 10 (71\%). Only 2 out of the 9 teams scored below average, indicating they may have benefited from an intervention by an instructor.

The TTX provided the teams with 7 possible tools. The team that reached the fewest milestones also used the least amount of tools (6 times in total compared to the overall team average of 20 occurrences of tool usage). However, the team that reached the second smallest number of milestones had the second largest number of tool usages. While other explanations are possible, this may indicate that rare tool usage is associated with low exercise completion, but frequent tool usage does not necessarily imply success (the tools might not be used efficiently).

\subsubsection{Run 3}

In order to improve the granularity of capturing teams' actions, we added 8 additional milestones to the TTX. When looking just at the 14 original milestones, the teams reached between 4 and 11 of them, with an average of 8 (57\%). Compared to Run 2, this lower ratio indicates that if an instructor provided a post-exercise debrief, it might be a valuable learning experience for all trainees. This debrief would inform the trainees about the additional actions that could have been made but were missed.

Overall, the first reached milestone was to visit the compromised website. The first team that reached this milestone did so in around 8 minutes from the TTX start. This milestone was also the fastest to reach across all teams, in 15.5 minutes on average. However, the slowest team to achieve this milestone took 30 minutes. Instead, this team focused on four other milestones before, prioritizing different aspects of the TTX compared to the vast majority of teams. This can be an interesting observation for the instructor, showing possibly alternative approaches to solving the in-exercise problems. Regardless, this team took rather long to reach their first milestone, which suggests they may have benefited from a hint or intervention.

The milestones that took the longest to complete encompassed the communication with the simulated employees. Four teams took a little more than an hour to address their stakeholders, and this step was completely overlooked by five teams. Although this skill is non-technical, it is still an important part of the cyber incident responders' work role. Therefore, instructors can use these insights from the TTX data to remind the learners about this responsibility or revise the course content in this aspect.

The TTX provided the teams with 11 possible tools (additional 4 compared to Run 2). A team used a tool between 10 and 46 times throughout the entire TTX, with 31 uses on average (including repeated uses of the same tool). The approaches of individual teams differed vastly: different teams used certain tools more often and (almost) ignored other tools. For example, all teams used the tool to block traffic incoming \textit{from} a certain IP address, but only two-thirds of the teams blocked traffic outgoing \textit{to} the compromised website.

Finally, the improvement of \ttxplatformshort\ for Run 3 enabled to evaluate the syntactical correctness of tool usage. When looking at these data, all tools have much more correct rather than incorrect invocations, showing that all trainees understood the tools' interfaces. However, the DNS lookup tool has substantially higher percentage of erroneous applications compared to other tools. Within the errors that were not simply typos, there might have been confusion among the trainees that could be addressed by the instructor.

Looking at the teams' written communication, they engaged in 6 email threads on average. The team that communicated the most (9 threads) reached the most milestones, and vice versa, the team that communicated the least (3 threads) reached almost the least number of milestones. This provides a teaching opportunity if the instructor compares the differences between the teams, showing that active communication is crucial while resolving a crisis.

\subsubsection{Summary}

The automatic collection and analysis of data provided by the \ttxplatformshort\ equips instructors and researchers with valuable insights that would be difficult to obtain otherwise, especially in the traditional pen-and-paper TTX format. By adding more granularity to the milestones and enhancing the platform's logging capabilities, we were able to observe deeper insights in Run 3 compared to Run~2. These include difficult milestones and errors in tool usage.

\subsection{Trainees' Learning Experience in \ttxplatformshort}

To complement the analysis of exercise logs, we present the results of a post-exercise survey administered to all trainees after Run~3.

In the overall evaluation, 35 out of 36 learners considered the TTX scenario realistic. A majority, 29 out of 36, found the TTX beneficial for practical applications because it improved their understanding of incident handling. One participant stated, \enquote{\textit{At the beginning, we were quite lost. There is just so much difference between having a specified incident handling task and having to figure out everything by yourself.}} Additionally, 31 participants expressed satisfaction with the ease of use of \ttxplatformshort\ to facilitate the exercise.

Our survey unveiled three pivotal insights for refining future exercises. Primarily, we encountered challenges in effectively communicating which in-exercise email addresses are trustworthy. Consequently, specific teams refrained from accessing some exercise emails, deeming them potentially malicious.

Secondly, trainees would like \ttxplatformshort's email feature to resemble familiar interfaces more. The current version can send and receive emails, but the trainees expected more features, like auto-saving drafts or showing emails in threads. The absence of such features led to communication delays, influencing the learning experience.

Finally, some teams got stuck in various stages of the scenario. Instructors were briefed to assist by sending exercise emails to guide these teams. However, this approach proved challenging as instructors struggled to identify the right moments for intervention. Given the continuous team discussions, instructors found it hard to determine when it was appropriate to influence the discussion.

\subsection{Instructors' Teaching Experience in \ttxplatformshort}

During a focus group discussion hosted with the instructors after Run 3, the instructors observed that enhancing \ttxplatformshort\ led to improving the following two key aspects in the teaching practice:
\begin{itemize}
    \item \textit{Reliability}: When using shared documents in Run 1, students sometimes accidentally rewrote their past conversation, and instructors got confused when working with multiple teams. A dedicated learning environment eliminates these errors.
    \item \textit{Involvement}: Run 1 and 2 had fewer teams, almost exclusively with 4 people. With the improvement of \ttxplatformshort\ for Run~3, we were able to have more teams, almost exclusively with 3~people. This means that a single student got more opportunities to speak in the team and to be actively involved in the decision-making, improving their individual experience.
\end{itemize}

\section{Conclusion}
\label{subsec:conclusion-materials}

Tabletop exercises are a promising method for innovating computing courses. They enable students to exercise collaborative problem-solving in the context of cybersecurity, IT governance, and other domains of applied informatics.

Introducing the \ttxplatform, a dedicated learning environment for TTXs, alleviates many challenges that instructors face. For example, having a platform to automate repetitive tasks, such as providing injects or outputs of tools, enables instructors to focus on the exercise facilitation. The automation capabilities of the learning environment also enable further educational research.

We release the \ttxplatformshort\ as open-source software with an example exercise at \url{https://inject.muni.cz}. The research data, Python scripts for data processing, and complete results are also available at \url{https://gitlab.fi.muni.cz/inject/papers/2024-iticse-from-paper-to-platform}.

\subsection{Open Research Challenges}

Currently, it is difficult to identify the right moments for intervention during TTXs. A key challenge is how to determine when a team would benefit from a hint, using insights from exercise data. For example, measuring expected time to reach a milestone can imply how long the instructor should wait before giving a team a hint. This would help teams to navigate through scenario challenges. 

Another limitation is that \ttxplatformshort\ does not yet support instructors in quickly reacting to expected trainee responses. Therefore, future work can explore machine learning and natural language processing techniques to evaluate the similarity in the responses to injects between different teams. Then, the platform can provide instructors with pre-defined responses that would suit the trainees' inputs.

Finally, future work should use the data to measure team performance~\cite{Amon2019}, evaluate students' achievement of learning objectives, and experimentally determine the effect of \ttxplatformshort\ on skill acquisition.

\begin{acks}
This research was supported by the Open Calls for Security Research 2023--2029 (OPSEC) program granted by the Ministry of the Interior of the Czech Republic under No. VK01030007 -- \textit{Intelligent Tools for Planning, Conducting, and Evaluating Tabletop Exercises}.
\end{acks}

\bibliographystyle{ACM-Reference-Format}
\balance
\bibliography{references}

\end{document}